\documentclass[aps,preprint]{revtex4-1}
\usepackage{graphicx}
\usepackage{amsmath}
\usepackage{makeidx}
\usepackage{amsfonts}
\usepackage{amssymb}

\begin{document}

\title{Loop corrections for hard spheres in Hamming space} 

\author{Abolfazl Ramezanpour}
\email{aramezanpour@gmail.com}
\affiliation{Department of Physics, College of Science, Shiraz University, Shiraz 71454, Iran}
\affiliation{Medical Systems Biophysics and Bioengineering, Leiden Academic Centre for Drug Research, Faculty of Science, Leiden University, Leiden, The Netherlands}	
\author{Saman Moghimi-Araghi}
\email{samanimi@sharif.edu}
\affiliation{Physics Department, Sharif University of Technology, P.O. Box 11155-9161 Tehran, Iran}

\date{\today}

\begin{abstract}
We begin with an exact expression for the entropy of a system of hard spheres within the Hamming space. This entropy relies on probability marginals, which are determined by an extended set of Belief Propagation (BP) equations. The BP probability marginals are functions of auxiliary variables which are introduced to model the effects of loopy interactions on a tree-structured interaction graph. We explore various reasonable and approximate probability distributions, ensuring they align with the exact solutions of the BP equations. Our approach is based on an ansatz of (in)homogeneous cavity marginals respecting the permutation symmetry of the problem. Through thorough analysis, we aim to minimize errors in the BP equations. Our findings support the conjecture that the maximum packing density asymptotically conforms to the lower bound proposed by Gilbert and Varshamov, further validated by the solution of the loopy BP equations.
\end{abstract}

\maketitle

\section{Introduction}\label{S0}
Finding the maximum packing density of hard spheres is a challenging problem in physics and coding theory, especially in high dimensions \cite{CS-book-1999,Z-book-2008,PZ-revmp-2010,TS-revmp-2010}. For example, exact solutions are only known for very small and specific dimensions like 1, 2, 3, 8, and 24\cite{F-mathz-1940,H-anmath-2005,V-anmath-2017,CKMRV-2017}. Currently, there's a significant gap between the best lower and upper bounds on the maximum packing density in very high dimensions \cite{MRRW-ieee-1977,B-mathres-1992,DL-ieee-1998,S-jcomb-2001,JV-ieee-2004,SST-jmat-2008,C-geo-2002,KLV-mathres-2004,ACHT-jhep-2020,CT-mathc-2022}. One approach to tackle this problem is to start with a crystalline structure made of elementary cells with a flexible basis. These cells form a periodic density distribution of the spheres, determined by the shape of the elementary cell and the arrangement of the spheres in the basis. By adjusting these parameters, we can find an upper bound for the maximum packing density that satisfies certain equations for the spatial density distribution.

On the other hand, the problem can be understood as mapping to a physical system of interacting particles, wherein the relevant packing configurations, weighted by the Boltzmann factor, are examined as the number density of hard spheres increases. Mean-field approximations are commonly employed to construct a phase diagram of the system, particularly in very high (infinite) dimensions where this approximation is expected to perform well if handled with care \cite{PS-jstat-1999,PS-pre-2000,PZ-jstat-2006,CKPUZ-anrevcm-2017}.

In the following, we will focus on the packing problem within the binary Hamming space. At zero temperature, this translates into a constraint satisfaction problem where any two spheres cannot overlap. The Bethe approximation (cavity method) has been utilized to estimate the maximum packing density in the Hamming space. Nevertheless, the degree to which these results accurately reflect the behaviour as the dimensionality approaches infinity in this fully-connected model with numerous interconnected interactions remains uncertain.

In this study we try a different approach by first introducing some auxiliary variables to map the original problem to an extended problem with a tree interaction graph where the Bethe approximation is exact \cite{R-pre-2013}. Then we write an exact expression for the entropy in terms of the (unique) solution to the Bethe, or Belief Propagation (BP), equations \cite{MP-physc-2001,KFL-ieee-2001,YFW-artint-2003,MM-book-2009}. To solve these equations and estimate the system's entropy, we make reasonable and manageable approximations.

The structure of the paper is as follows. We begin with a more precise statement of the problem and a summary of the results.  In Sec. \ref{S1} we present the mapping to a tree interaction graph and write the associated Bethe equations and the entropy.   
Section \ref{S2} is devoted to finding approximate BP solutions, starting with a naive homogeneous (liquid) solution and ending with inhomogeneous solutions which respect the permutation symmetry of the problem. The concluding remarks and some details of the results are given in Sec. \ref{S3} and Appendices \ref{app1}-\ref{app2}, respectively.

\subsection{The problem statement}\label{S01}
Consider the binary Hamming space in $n$ dimensions. In other words, an $n$-dimensional discrete space where each dimension can take either zero or one. We want to place $N$ hard spheres of diameter $d$ on the points of this lattice in way that the spheres do not overlap. However, if they have no overlap, they are considered completely non-interacting. Placing a hard sphere at any point $\vec{\sigma}_i\in (0,1)^n$ produces a forbidden or occupied space which is the set of discrete points within the ball of radius $d$ at which the other hard spheres cannot be located. This volume is given by
\begin{align}
V_d(\vec{\sigma}_i)=\{\vec{\sigma}_j:D_{ij}<d\},
\end{align}
as the distance between any two points in Hamming space is equal to 
\begin{align}
D_{ij}=\sum_{a=1}^n(\sigma_i^a-\sigma_j^a)^2.
\end{align}
And if the coordinates of any two points are not identical in exactly $q$ dimensions, their distance from each other is equal to $D_{ij}=q$. Thus, the number of inaccessible points is equal to 
\begin{align}
V_d=|V_d(\vec{\sigma}_i)|=\sum_{l=0}^{d-1}C(l:n),
\end{align}
where $C(l:n)=n!/(l!(n-l)!)$. Note that this number differs from the volume of a spherical shell with radius $d$.

Let's write the partition function of this system. Since the spheres are considered non-interacting, the partition function of the system is just the number of possible configurations for the packing. Any arbitrary configuration of these $N$ spheres in the Hamming space is represented by 
\begin{align}
\vec{\boldsymbol{\sigma}}=\{\vec{\sigma}_i \in (0,1)^n: i=1,\cdots,N\}.
\end{align}
and thus, the partition function is given by 
\begin{align}
Z_n(N,d)=\sum_{\vec{\boldsymbol\sigma}} \prod_{i<j} \mathbb{I}(D_{ij}\ge d)=e^{S_n(N,d)}.
\end{align}
where the indicator function $\mathbb{I}(C)=1$ if constraint $C$ is satisfied, otherwise $\mathbb{I}(C)=0$.
As mentioned, the partition function represents the number of possible configurations for packing. Naturally, in a space of a certain size, if we increase the number of spheres, we reach a point where no valid configuration can be found and the spheres would surely have overlaps. We call the maximum number of spheres for which a valid configuration is found $N_{\text{max}}$. In other words, we must have 
\begin{align}
Z_n(N_{max},d)> 0,\hskip1cm Z_n(N_{max}+1,d)= 0. 
\end{align}
This quantity and other properties of the system depend on the space dimension $n$ and the diameter considered for the spheres $d$. Increasing $n$ enlarges the space, while increasing \(d\) reduces the number of states. We are interested in the case where $d, n \rightarrow \infty$ while their ratio remains constant and equal to $\delta$.

Within the Bethe approximation and assuming the replica symmetry \cite{MM-book-2009,baldassi-polito-2009}, the entropy $S_n(N,d)=\ln Z_n(N,d)$ is estimated by
\begin{align}
S_n^{LBP}(N,d)=\sum_{i}\Delta S_i-\sum_{i<j}\Delta S_{ij}. 
\end{align}
The entropy contributions of the nodes (spheres) $\Delta S_i$ and edges (interactions) $\Delta S_{ij}$ depend on the solutions to the Bethe equations
\begin{align}
\eta_{i\to j}(\vec{\sigma}_i) \propto \prod_{k\ne i,j}\left(\sum_{\vec{\sigma}_k} \mathbb{I}(D_{ik}\ge d)\eta_{k\to i}(\vec{\sigma}_k)\right).
\end{align}
Here the cavity marginal $\eta_{i\to j}(\vec{\sigma}_i)$ is the probability of finding sphere $i$ in position $\vec{\sigma}_i$ in the absence interaction with sphere $j$. Given a fixed point of the above equation, one computes the entropy contributions
\begin{align}
e^{\Delta S_i} &= \sum_{\vec{\sigma}_i}\prod_{j\ne i}\left(\sum_{\vec{\sigma}_j} \mathbb{I}(D_{ij}\ge d)\eta_{j\to i}(\vec{\sigma}_j)\right),\\
e^{\Delta S_{ij}} &= \sum_{\vec{\sigma}_i,\vec{\sigma}_j}\mathbb{I}(D_{ij}\ge d)\eta_{i\to j}(\vec{\sigma}_i)\eta_{j\to i}(\vec{\sigma}_j).
\end{align}
 
For a tree interaction graph, the Bethe equations with the unique solution $\eta_{i\to j}(\vec{\sigma}_i)=1/2^n$ provide the exact entropy:
\begin{align}
S_n^{T}(N,d)=N\ln(2^n)+(N-1)\ln (1-\frac{V_d}{2^n}),
\end{align}
which is greater than zero for any $d<n$, irrespective of the tree structure.

The entropy obtained by the loopy Belief Propagation (LBP) equations with a homogeneous or liquid solution which respects the translation symmetry of the problem $(\eta_{i\to j}(\vec{\sigma}_i)=1/2^n)$ is:
\begin{align}\label{Sbethe}
S_n^{LBP}(N,d)=N\ln(2^n)+\frac{N(N-1)}{2}\ln (1-\frac{V_d}{2^n}).
\end{align}
The maximum $N$ here is given by
\begin{align}\label{Nbethe}
N_{max}^{LBP}=1-\frac{2\ln 2^n}{\ln (1-v_d)}\to (2\ln 2)\frac{n}{v_d},
\end{align}
where $v_d=\frac{V_d}{2^n}$. 

Note that the Gilbert-Varshamov (GV) lower bound states that
\begin{align}\label{GV}
N_{max}\ge \frac{2^n}{V_d}=\frac{1}{v_d}=N_{max}^{GV}.
\end{align}
It seems that even other solutions to the loopy Bethe equations do not result in an exponentially larger maximum number of spheres than the GV lower bound \cite{RZ-pre-2012}.

\subsection{Summary of the results}\label{S02}
The main results of this paper are listed below:
\begin{itemize}

\item The interaction graph of spheres is partitioned into a tree interaction graph and a set of induced loopy interactions. The effect of loopy interactions is represented by messages passing along the edges of the tree graph. We formulate the BP equations with these auxiliary variables in an extended space to obtain an exact expression for the entropy of the packing problem in the Hamming space.

\item A naïve ansatz for the BP probability marginals reproduces the entropy, previously obtained by the loopy Bethe equations (assuming replica symmetry or one step of replica symmetry breaking). This asymptotically coincides with the GV lower bound for the maximum number of spheres.

\item We observe that the aforementioned naïve ansatz asymptotically satisfies (“on average”) the extended BP equations. Similar results are obtained with other reasonable candidates for homogeneous (liquid) BP marginals, which are expected to be closer to the exact solution of the extended BP equations.

\item Numerically, we observe that an initially inhomogeneous solution to the BP equations approaches the homogeneous one as the number of spheres increases, starting from two neighboring spheres localized in the Hamming space. However, this does not conclusively prove that there is no packing density that asymptotically exceeds the GV lower bound.

\end{itemize}

\section{An exact tree representation}\label{S1}
\begin{figure}
\includegraphics[width=12cm]{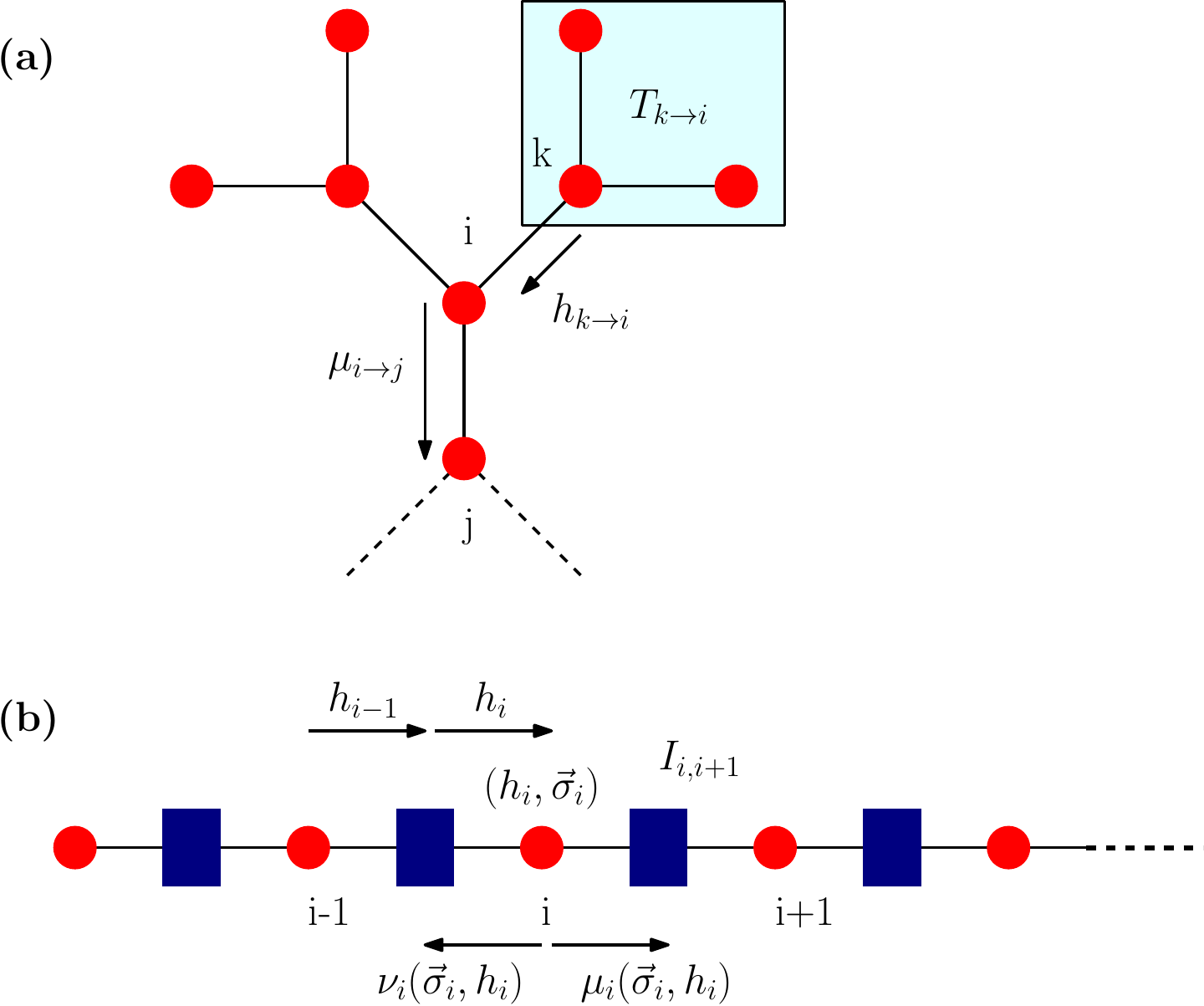} 
\caption{Illustration of a factor graph with the variables, constraints, and messages. The original variables are the sphere positions $\vec{\sigma}_i$ besides the auxiliary variables (internal messages) $h_{i\to j}$ which are to represent the effect of loopy interactions. The cavity probability marginals (external BP messages) of the extended set of variables are $\mu_{i\to j}(\vec{\sigma}_i,h_{i\to j})$.  For simplicity, in this study we shall work with a chain factor graph (Panel (b)).}\label{fig1}
\end{figure}

Consider a connected tree graph $T$ of $N$ nodes and the associated local branches or cavity trees $T_{i\to j}$. More precisely, $T_{i\to j}$ is the subgraph that is obtained after removing edge $(ij)$ with root node $i$. This can recursively be defined as follows
\begin{align}
T_{i\to j}=i \cup_{k\in \partial i\setminus j} T_{k\to i}.
\end{align}
Here $\partial i$ denotes the set of neighbors of node $i$ in the graph. Figure \ref{fig1} displays such a tree interaction graph.  
The size of such cavity tree is denoted by
\begin{align}
N_{i\to j}=|T_{i\to j}|.
\end{align}

For a given configuration $\vec{\boldsymbol{\sigma}}$, let us define the cavity messages $h_{i\to j}$ to represent the space occupied by all the spheres in the cavity tree $T_{i\to j}\setminus i$, that is
\begin{align}
h_{i\to j}=\cup_{k\in T_{i\to j}\setminus i} V_d(\vec{\sigma}_k).
\end{align}
Or, in terms of the other incoming messages 
\begin{align}
h_{i\to j}=\cup_{k\in \partial i \setminus j} \left( h_{k\to i} \cup V_d(\vec{\sigma}_k) \right).
\end{align}
Equivalently, the messages $h_{i\to j}$ can be considered as the positions of all the spheres in $T_{i\to j}$ except sphere $i$.
These message are defined to have access to the positions of all the spheres locally at each node of the tree graph $T$.

Now with the extended set of variables $\vec{\boldsymbol\sigma}$ and $\mathbf{h}=\{h_{i\to j}: i=1,\dots,N, j\in \partial i\}$, we rewrite the partition function of the sphere packing problem,
\begin{multline}
Z_n(N,d)=\sum_{\vec{\boldsymbol\sigma}} \prod_{(ij)\in T} \mathbb{I}(D_{ij}\ge d) \\
\times \left(\sum_{\mathbf{h}}\prod_i \prod_{j\in \partial i}\mathbb{I}(V_d(\vec{\sigma}_i)\cap h_{j\to i}=\emptyset)\mathbb{I}(h_{i\to j}=\cup_{k\in \partial i\setminus j} h_{k\to i}\cup V_d(\vec{\sigma}_k))\right).
\end{multline}
The constraints are to ensure that the messages $h_{i\to j}$ satisfy the necessary equations and at the same time any two spheres have a distance larger than or equal to $d$. Note that there is only one solution to the constraints on the messages $h_{i\to j}$ for a given configuration $\vec{\boldsymbol\sigma}$ of the spheres.    

The Bethe equations for the above problem are recursive equations for the cavity probability distributions $\mu_{i\to j}(\vec{\sigma_i},h_{i\to j}:\vec{\sigma_j},h_{j\to i})$ along the directed edges of the tree graph. Here $\mu_{i\to j}(\vec{\sigma_i},h_{i\to j}:\vec{\sigma_j},h_{j\to i})$ is the probability of having $(\vec{\sigma_i},h_{i\to j})$ as the position of sphere $i$ and the message $h_{i\to j}$ conditioned on the values of $(\vec{\sigma_j},h_{j\to i})$. In a tree graph, the incoming variables are independent of each other, therefore \cite{MM-book-2009}, 
\begin{multline}
\mu_{i\to j}(\vec{\sigma_i},h_{i\to j}:\vec{\sigma_j},h_{j\to i}) \propto \mathbb{I}(\vec{\sigma_i},h_{j\to i})\\
\times \sum_{\{\vec{\sigma}_k,h_{k\to i}:k\in \partial i\setminus j\}}\mathbb{I}(h_{i\to j})\prod_{k\in \partial i\setminus j}\left[\mathbb{I}(D_{ik}) \mathbb{I}(\vec{\sigma_i},h_{k\to i})\mathbb{I}(h_{i\to k})\mu_{k\to i}(\vec{\sigma_k},h_{k\to i}:\vec{\sigma_i},h_{i\to k})\right],
\end{multline}
where for brevity's sake, we defined
\begin{align}
\mathbb{I}(\vec{\sigma_i},h_{k\to i}) &=\mathbb{I}(V_d(\vec{\sigma}_i)\cap h_{k\to i}=\emptyset),\\
\mathbb{I}(D_{ik}) &=\mathbb{I}(D_{ik}\ge d),\\
\mathbb{I}(h_{i\to k}) &=\mathbb{I}(h_{i\to k}=\cup_{j\in \partial i\setminus k} h_{j\to i}\cup V_d(\vec{\sigma}_j)).
\end{align}
In the following, the messages $h_{i\to j}$ are called internal messages and the cavity probabilities $\mu_{i\to j}$ are called the external BP messages. In practice, the BP equations are solved by iteration in a random sequential way starting with initial BP messages $\mu_{i\to j}$. Here the tree structure of the interaction graph insures that there a unique solution to the above equations.

It is straightforward to start from the partition function $Z_n(N,d)$ of the tree interaction graph $T$ and relate the free entropy $S_n(N,d)$ to the BP cavity marginals \cite{baldassi-polito-2009}. This can be written in term of the variables (nodes) and interactions (edges) contributions to the entropy 
\begin{align}
S_n(N,d)=\sum_i \Delta S_i-\sum_{(ij)\in T} \Delta S_{ij},
\end{align}
where
\begin{align}
e^{\Delta S_i}=\sum_{\vec{\sigma}_i} \sum_{\{\vec{\sigma}_j,h_{j\to i}:j\in \partial i\}}\prod_{j\in \partial i}\left[\mathbb{I}(D_{ij}) \mathbb{I}(\vec{\sigma_i},h_{j\to i})\mathbb{I}(h_{i\to j})\mu_{j\to i}(\vec{\sigma_j},h_{j\to i}:\vec{\sigma_i},h_{i\to j})\right],
\end{align}
and
\begin{align}
e^{\Delta S_{ij}}= \sum_{\vec{\sigma}_i,h_{i\to j},\vec{\sigma}_j,h_{j\to i}} \mathbb{I}(D_{ij}) \mu_{i\to j}(\vec{\sigma_i},h_{i\to j}:\vec{\sigma_j},h_{j\to i})\mu_{j\to i}(\vec{\sigma_j},h_{j\to i}:\vec{\sigma_i},h_{i\to j}).
\end{align}

In the following we shall work with a tree structure $T$ which is represented by a chain of interacting spheres.
This allows us to simplify the BP equations and obtain simpler expressions for the BP cavity marginals and the entropy.

\subsection{Chain representation}\label{S11}
Here we assume that the spheres are arranged in a chain from $i=1,\dots,N$. Besides the $\vec{\sigma}_i$ we need the messages $h_i$ passing from left to right. Thus on each edge $(i,i+1)$ we have three constraints: $D_{i,i+1}\ge d$, $h_{i+1}=h_i\cup  V_d(\vec{\sigma}_i)$, and  $h_i \cap V_d(\vec{\sigma}_{i+1})=\emptyset$. Let the indicator function $\mathbb{I}_{i,i+1}$ represent these constraints. Then, the left-to-right BP messages $\mu_{i}(\vec{\sigma}_i,h_i)$ and right-to-left BP messages $\nu_{i}(\vec{\sigma}_i,h_i)$ are given by the following BP equation,
\begin{align}
\mu_{i}(\vec{\sigma}_i,h_i) \propto \sum_{\vec{\sigma}_{i-1},h_{i-1}}\mathbb{I}_{i-1,i} \mu_{i-1}(\vec{\sigma}_{i-1},h_{i-1}),\\
\nu_{i}(\vec{\sigma}_i,h_i) \propto \sum_{\vec{\sigma}_{i+1},h_{i+1}}\mathbb{I}_{i,i+1} \nu_{i+1}(\vec{\sigma}_{i+1},h_{i+1}).
\end{align}
The boundary messages are
\begin{align}
\mu_{1}(\vec{\sigma},h_1) &= \frac{1}{2^n}\delta_{h,\emptyset},\\
\nu_{N}(\vec{\sigma},h_N) &= \frac{1}{2^n}\frac{1}{2^{n(N-1)}}.
\end{align}
Here we are considering $h_i$ as the set of positions of spheres $j=1,\dots,i-1$.
In this way, the right-to-left messages are uniform distributions. Also away from the boundary
\begin{align}
\nu_{i}(\vec{\sigma},h) = \frac{1}{2^n}\frac{1}{2^{n(i-1)}}.
\end{align}

The entropy within the chain representation reads
\begin{align}
S_n(N,d)=\sum_{i=1}^N \Delta S_i-\sum_{i=1}^{N-1} \Delta S_{i,i+1},
\end{align}
where now the node contribution is
\begin{align}\label{dsi-chain}
e^{\Delta S_i}=\sum_{\vec{\sigma}_i,h_i} \left(\sum_{\vec{\sigma}_{i-1},h_{i-1}}\mathbb{I}_{i-1,i} \mu_{i-1}(\vec{\sigma}_{i-1},h_{i-1})\right)\left(\sum_{\vec{\sigma}_{i+1},h_{i+1}}\mathbb{I}_{i,i+1} \nu_{i+1}(\vec{\sigma}_{i+1},h_{i+1})\right),
\end{align}
and the edge contribution  is given by
\begin{align}\label{dsij-chain}
e^{\Delta S_{i,i+1}}= \sum_{\vec{\sigma}_i,h_i,\vec{\sigma}_{i+1},h_{i+1}} \mathbb{I}_{i,i+1} \mu_{i}(\vec{\sigma}_i,h_i)\nu_{i+1}(\vec{\sigma}_{i+1},h_{i+1}).
\end{align}

In the following we shall work with the chain representation.

\section{Approximate solutions of the Bethe equations}\label{S2}

\subsection{A naive approximation of the BP messages}\label{S21}
Let us write the BP probability marginals $\mu_{i}(\vec{\sigma}_i,h_i)$ in term of probability of position of sphere $i$, $p_i(\vec{\sigma}_i)$ and a conditional probability $q_i(h_i:\vec{\sigma}_i)$, 
\begin{align}
\mu_{i}(\vec{\sigma}_i,h_i)= p_i(\vec{\sigma}_i) q_i(h_i:\vec{\sigma}_i),
\end{align}
with $h_i=\{\vec{\sigma}_{j=1,\dots,i-1}\}$.
Then, by symmetry we take a uniform distribution for $\vec{\sigma}_i$,
\begin{align}
p_{i}(\vec{\sigma}_i)=\frac{1}{2^n}.
\end{align}
We also assume that the probability distribution of the other spheres is factorized and uniform
\begin{align}
q_{i}(h_{i}:\vec{\sigma}_i)=\prod_{j=1}^{i-1}\frac{\mathbb{I}(D_{i,j}\ge d)}{2^n-V_d}.
\end{align}

\begin{figure}
\includegraphics[width=12cm]{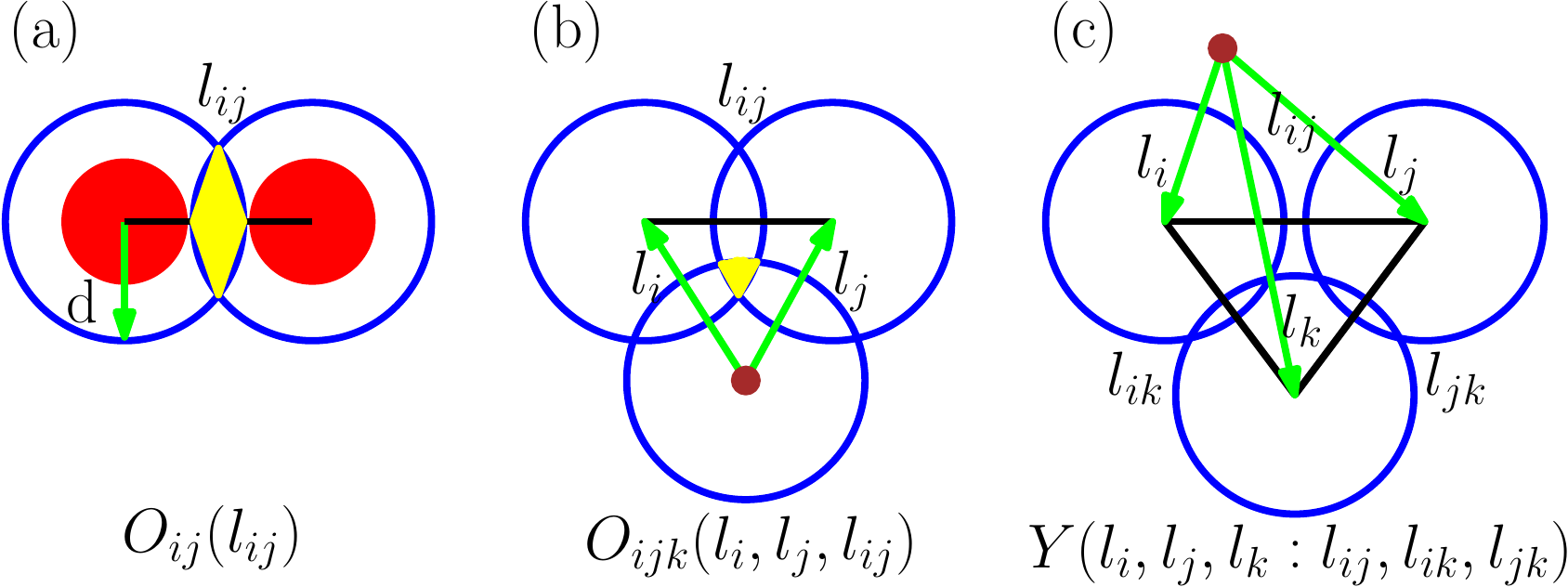} 
\caption{An illustration of the geometrical measures defined by a few hard spheres of diameter $d$. The solid discs display the hard spheres. The empty discs show the space which can not be occupied by the center of another hard sphere. Variable $l_{ij}$ denotes the distance between two spheres $i$ and $j$ with overlap $O_{ij}(l_{ij})$. Distance of a point in space from sphere $i$ is denoted by $l_i$.}\label{fig2}
\end{figure}

In this way, for the node contributions to the entropy from Eq. \ref{dsi-chain} we obtain
\begin{multline}
e^{\Delta S_i}=\frac{1}{2^{n+ni}}\sum_{l_1,l_2,l_{12}=d}^n C(l_{12}:n)
\Omega(l_1,l_2:l_{12})\\
\times \left(1-\frac{2V_d-O_{01}(l_1)-O_{02}(l_2)-O_{12}(l_{12})+2O_{012}(l_1,l_2,l_{12})}{2^n-V_d}\right)^{i-2}.
\end{multline}
The other terms in the entropy come from the interactions along the chain and are obtained from Eq. \ref{dsij-chain}
\begin{align}
e^{\Delta S_{i,i+1}}= \frac{1}{2^{n+ni}}\sum_{l=d}^n C(l:n)
\left(1-\frac{V_d-O_{01}(l)}{2^n-V_d}\right)^{i-1}.
\end{align}
Here $O_{ij}(l)$ is the overlap of two spheres of radius $d$ at distance $l$. $O_{ijk}(l_i,l_j,l_{ij})$ is the overlap of three spheres of radius $d$ when $(i,j)$ have distance $l_{ij}$ and the other sphere is at distances $l_i,l_j$ form $i$ and $j$, respectively.
The function $\Omega(l_i,l_j:l_{ij})$ is the number of possible points for the third sphere given the distances. See Fig. \ref{fig2} for a schematic representation of these quantities.

Let us take the limit $n,d\to \infty$ with $\delta=d/n$ finite. All distances are scaled with $n$, for instance $r_{ij}=l_{ij}/n$. Then $v_d=V_d/2^n=e^{-n[\ln(2)-H(\delta)]}$, with $H(\delta)=-\delta \ln(\delta)-(1-\delta) \ln(1-\delta)$ for the binary entropy, is an exponentially small quantity. In Appendix \ref{app1} we obtain the asymptotic entropy contributions of the nodes and edges in the total entropy,  
\begin{align}
\Delta S_i &\simeq nN\left(\frac{3}{2}\ln 2-\tilde{v}_d\right),\\
\Delta S_{i,i+1} &\simeq nN\frac{1}{2}\left(\ln 2-\tilde{v}_d\right),\\
S^{naive} &=\Delta S_{i}-\Delta S_{i,i+1}\simeq nN\left(\ln 2-\frac{1}{2}\tilde{v}_d\right).
\end{align}
where we defined $\tilde{v}_d=Nv_d/n$. Therefore, the entropy vanishes at
\begin{align}
N_{max}^{naive}=(2\ln 2)\frac{n}{v_d},
\end{align}
at the same point that the loopy BP entropy $S_n^{LBP}(N,d)$ in Eq. \ref{Sbethe} goes to zero.

\subsection{Beyond the naive approximation}\label{S22}
Note that the probability distribution of the spheres is invariant under the permutation of the spheres. Therefore, by de Finetti theorem \cite{KLC-pra-2013} the probabilities $q_{i}(h_{i}:\vec{\sigma}_i)$ can well be approximated by a convex combination of factorized distributions,   
\begin{align}
q_{i}(h_{i}:\vec{\sigma}_i)=\sum_{\alpha}c_i(\alpha)\prod_{j=1}^{i-1}f_{i,j}^{\alpha}(\vec{\sigma}_j:\vec{\sigma}_i),
\end{align}
where
\begin{align}
\sum_{\alpha}c_i(\alpha) &=1,\\
\sum_{\vec{\sigma}_j}f_{i,j}^{\alpha}(\vec{\sigma}_j:\vec{\sigma}_i) &=1.
\end{align}
Recall that $h_i$ is equivalent to the set of positions $\{\vec{\sigma}_{j=1,\dots,i-1}\}$. 
Let us for simplicity consider only one of the factorized terms. Then the BP equations are
\begin{align}
p_i(\vec{\sigma}_i)\prod_{j=1}^{i-1}f_{ij}(\vec{\sigma}_j:\vec{\sigma}_i)
\propto \sum_{\vec{\sigma}_{i-1},h_{i-1}}\mathbb{I}_{i-1,i} p_{i-1}(\vec{\sigma}_{i-1})\prod_{j=1}^{i-2}f_{i-1,j}(\vec{\sigma}_j:\vec{\sigma}_{i-1}).
\end{align}
It means that 
\begin{multline}
p_i(\vec{\sigma}_i)f_{i,i-1}(\vec{\sigma}_{i-1}:\vec{\sigma}_i)\prod_{j=1}^{i-2}f_{i,j}(\vec{\sigma}_j:\vec{\sigma}_i)\\
= \frac{p_{i-1}(\vec{\sigma}_{i-1})}{z_i}\mathbb{I}(D_{i,i-1}\ge d)\prod_{j=1}^{i-2}\mathbb{I}(D_{i,j}\ge d)f_{i-1,j}(\vec{\sigma}_j:\vec{\sigma}_{i-1}),\\
\end{multline}
where the normalization factor $z_i$ is
\begin{align}\label{zi}
z_i= \sum_{\vec{\sigma}_i,h_i} p_{i-1}(\vec{\sigma}_{i-1})\mathbb{I}(D_{i,i-1}\ge d)\prod_{j=1}^{i-2}\mathbb{I}(D_{i,j}\ge d)f_{i-1,j}(\vec{\sigma}_j:\vec{\sigma}_{i-1}).
\end{align}
Taking $p_i(\vec{\sigma}_i)=p_{i-1}(\vec{\sigma}_{i-1})=1/2^n$, the above equation suggests the following solution
\begin{align}
f_{i,j}(\vec{\sigma}_j:\vec{\sigma}_i)=\frac{\mathbb{I}(D_{i,j}\ge d)}{2^n-V_d},
\end{align}
for $j=1,\dots,i-2$, and
\begin{align}
f_{i,i-1}(\vec{\sigma}_{i-1}:\vec{\sigma}_i)=\mathbb{I}(D_{i,i-1}\ge d)g_{i,i-1}(\vec{\sigma}_{i-1}:\vec{\sigma}_i),
\end{align}
for $j=i-1$. Given the $\{f_{i-1,j}:j=1,\dots,i-2\}$, the function $g_{i,i-1}(\vec{\sigma}_{i-1}:\vec{\sigma}_i)$ should be chosen according to the following constraints,
\begin{align}
\sum_{\vec{\sigma}_{i-1}}g_{i,i-1}(\vec{\sigma}_{i-1}:\vec{\sigma}_i)\mathbb{I}(D_{i,i-1}\ge d) &=1,\\
g_{i,i-1}(\vec{\sigma}_{i-1}:\vec{\sigma}_i)
&=\frac{(2^n-V_d)^{i-2}}{z_i} \prod_{j=1}^{i-2}f_{i-1,j}(\vec{\sigma}_j:\vec{\sigma}_{i-1}).
\end{align}
The equations can not be true for any configuration. To overcome the problem of dependence on $\vec{\sigma}_j$, we can simply average over these variables using a uniform measure, leading to
\begin{multline}\label{gizi}
g_{i,i-1}(\vec{\sigma}_{i-1}:\vec{\sigma}_i)
= \frac{(2^n-V_d)^{i-2}}{z_i}\left\langle\prod_{j=1}^{i-2}f_{i-1,j}(\vec{\sigma}_j:\vec{\sigma}_{i-1})\right\rangle \\
=\frac{(1-v_d)^{i-2}}{z_i}\sum_{\vec{\sigma}_{i-2}} g_{i-1,i-2}(\vec{\sigma}_{i-2}:\vec{\sigma}_{i-1})\mathbb{I}(D_{i-1,i-2}\ge d)\\
=\frac{(1-v_d)^{i-2}}{z_i}.
\end{multline}
In this way $z_i=2^n(1-v_d)^{i-1}$, thus
\begin{align}
f_{i,i-1}(\vec{\sigma}_{i-1}:\vec{\sigma}_i)=\frac{\mathbb{I}(D_{i,i-1}\ge d)}{2^n-V_d}.
\end{align}
It means that all the $f_{i,j}$ for $j=1,\dots,i-1$ are given by the expression we used in the naive approximation of the messages. Moreover, the $f_{i,j}$ are the same function, as expected from the permutation symmetry of the problem.

Note that given the above $f_{i,j}$, according to Eq. \ref{zi},
\begin{align}
z_i=\frac{1}{(2^n-V_d)^{i-2}}\sum_{l_{12}=d}^nC(l_{12}:n)
\left(\sum_{l_1,l_2=d}^n
\Omega(l_1,l_2:l_{12})\right)^{i-2}.
\end{align}
For $n\to \infty$, after ignoring the exponentially small overlaps,
\begin{align}
z_i\simeq \frac{1}{(2^n-V_d)^{i-2}}(2^n-V_d)(2^n-2V_d)^{i-2}=2^n(1-v_d)(1-\frac{V_d}{2^n-V_d})^{i-2}\simeq 2^n(1-v_d)^{i-1},
\end{align}
which is consistent with the expression we obtained for this quantity after Eq. \ref{gizi}.

\subsection{Minimizing the BP errors}\label{S23}
In this subsection, we look for a better solution for $f_{i,j}(\vec{\sigma}_j:\vec{\sigma}_i)=\mathbb{I}(D_{i,j}\ge d)g_{i,j}(\vec{\sigma}_j:\vec{\sigma}_i)$ by minimizing the distance
\begin{align}
\mathcal{L}[\{g_{i,j=1,\dots,i-1}\}]=\sum_{\{\vec{\sigma}_{j=1,\dots,i-1}:D_{i,j}\ge d\}}\left( \prod_{j=1}^{i-1}g_{i,j}-\frac{1}{z_i}\prod_{j=1}^{i-2}\mathbb{I}(D_{i-1,j}\ge d)g_{i-1,j}\right)^2,
\end{align}
constrained with
\begin{align}
\sum_{\vec{\sigma}_j:D_{i,j}\ge d}g_{i,j}=1.
\end{align}
The stationary equations for the $g_{i,j}(\vec{\sigma}_j:\vec{\sigma}_i)$ result in
\begin{multline}
\lambda_{i,j}=g_{i,j}(\vec{\sigma}_j:\vec{\sigma}_i)\prod_{k=1:k\neq j}^{i-1}(\sum_{\vec{\sigma}_k:D_{i,k}\ge d}g_{i,k}^2)\\
-\frac{1}{z_i}\sum_{\vec{\sigma}_{i-1}:D_{i,i-1},D_{i-1,j}\ge d}g_{i,i-1}g_{i-1,j}\prod_{k=1:k\neq j}^{i-2}(\sum_{\vec{\sigma}_k:D_{i,k},D_{i-1,k}\ge d}g_{i,k}g_{i-1,k}),
\end{multline}
for $j=1,\dots,i-2$, and
\begin{align}\label{lii-1}
\lambda_{i,i-1}=g_{i,i-1}(\vec{\sigma}_{i-1}:\vec{\sigma}_i)\prod_{k=1}^{i-2}(\sum_{\vec{\sigma}_k:D_{i,k}\ge d}g_{i,k}^2)-\frac{1}{z_i}\prod_{k=1}^{i-2}(\sum_{\vec{\sigma}_k:D_{i,k}\ge d,D_{i-1,k}\ge d}g_{i-1,k}g_{i,k}).
\end{align}
The Lagrange multipliers $\lambda_{ij}$ are to ensure the normalization constraints.
Let us assume that the $g_{i,j}$ depend only on the Hamming distances $D_{i,j}$.
By symmetry we also assume that $g_{i,j}=g_i$ is the same for all $j=1,\dots,i-1$ and consider only the later equation for $g_{i,i-1}$. In this way, from Eq. \ref{lii-1} we get
\begin{align}
g_{i}(l)=\frac{\lambda_{i}+\frac{1}{z_i}\left(\sum_{l_1,l_2=d}^n \Omega(l_1,l_2:l)g_{i-1}(l_1)g_i(l_2)\right)^{i-2}}{\left(\sum_{l_1=d}^nC(l_1:n)g_i(l_1)^2\right)^{i-2}},
\end{align}
\begin{align}
z_i=\sum_{l=d}^nC(l:n)
\left(\sum_{l_1,l_2=d}^n\Omega(l_1,l_2:l)g_{i-1}(l_1) \right)^{i-2},
\end{align}
and for the Lagrange multiplier
\begin{multline}
\lambda_{i}=\frac{1}{2^n-V_d}\left(\sum_{l_1=d}^nC(l_1:n)g_i(l_1)^2\right)^{i-2}\\
-\frac{1}{z_i(2^n-V_d)}\sum_{l=d}^nC(l:n)\left(\sum_{l_1,l_2=d}^n \Omega(l_1,l_2:l)g_{i-1}(l_1)g_i(l_2)\right)^{i-2}.
\end{multline}
The above equations can be rewritten in a compact form as  
\begin{align}
g_i(l)=\frac{1}{2^n-V_d}+(X_i(l)-\langle X_i(l)\rangle),
\end{align}
where we defined
\begin{align}
X_i(l)=\frac{1}{z_i}\left( 
\frac{\sum_{l_1,l_2=d}^n \Omega(l_1,l_2:l)g_{i-1}(l_1)g_i(l_2)}{\sum_{l_1=d}^nC(l_1:n)g_i(l_1)^2}
\right)^{i-2},
\end{align}
and
\begin{align}
\langle X_i(l)\rangle=\frac{1}{2^n-V_d}\sum_{l=d}^nC(l:n)X_i(l).
\end{align}

Given the $g_{i-1}(l)$, the equations can be solved by iteration for $g_i(l)$, starting from $g_2(l)=1/(2^n-V_d)$.
We see that the trial solution $g_i(l)=1/(2^n-V_d)$, gives 
\begin{align}
\langle X_i(l)\rangle &=\frac{1}{2^n-V_d},
\end{align}
and
\begin{align}
X_i(l)=\frac{(1-v_d\frac{1-o_{ij}(l)}{1-v_d})^{i-2}}{\sum_{l=d}^nC(l:n)(1-v_d\frac{1-o_{ij}(l)}{1-v_d})^{i-2}},
\end{align}
where $v_d=V_d/2^n$ and $o_{ij}=O_{ij}/V_d$ are exponentially small quantities. Approximating $(1-o_{ij}(l))/(1-v_d)\approx 1$ is therefore consistent with the solution $g_i(l)=1/(2^n-V_d)$ obtained in the previous subsection.

\subsection{Considering the permutation symmetry}\label{S24}
Let us consider the general case where the conditional probability marginal in the BP messages is a convex combination of factorized forms
\begin{align}
q_{i}(h_{i}:\vec{\sigma}_i)=\sum_{\alpha}c_i(\alpha)\prod_{j=1}^{i-1}\left( \mathbb{I}(D_{i,j}\ge d)g_{i,j}^{\alpha}(\vec{\sigma}_j:\vec{\sigma}_i) \right).
\end{align}
Then we try to minimize the consistency error of the above trail solutions in the BP equations
\begin{multline}
\mathcal{L}[\{c_i^{\alpha},g_{i,j=1,\dots,i-1}^{\alpha}\}]\\
=\sum_{\{\vec{\sigma}_{j=1,\dots,i-1}:D_{i,j}\ge d\}}\left( \sum_{\alpha}c_i(\alpha)\prod_{j=1}^{i-1}g_{i,j}^{\alpha}-\frac{1}{z_i}\sum_{\alpha}c_{i-1}(\alpha)\prod_{j=1}^{i-2}\mathbb{I}(D_{i-1,j}\ge d)g_{i-1,j}^{\alpha}\right)^2,
\end{multline}
with respect to the variables $c_i(\alpha)$, and $g_{i,j}^{\alpha}$.

The stationary equations for the unknown variables lead to the following expressions
\begin{multline}
\lambda_i=\sum_{\alpha'}c_i(\alpha')\prod_{j=1}^{i-1}(\sum_{\vec{\sigma}_j:D_{i,j}\ge d}g_{i,j}^{\alpha}g_{i,j}^{\alpha'})\\
-\frac{1}{z_i}\sum_{\alpha'}c_{i-1}(\alpha')\sum_{\vec{\sigma}_{i-1}:D_{i,i-1}\ge d}g_{i,i-1}^{\alpha}\prod_{j=1}^{i-2}(\sum_{\vec{\sigma}_j:D_{i,j},D_{i-1,j}\ge d}g_{i,j}^{\alpha}g_{i-1,j}^{\alpha'}),
\end{multline}
and
\begin{multline}
\lambda_{ij}^{\alpha}=c_i(\alpha)\sum_{\alpha'}c_i(\alpha')g_{i,j}^{\alpha'}\prod_{k=1:k\neq j}^{i-1}(\sum_{\vec{\sigma}_k:D_{i,k}\ge d}g_{i,k}^{\alpha}g_{i,k}^{\alpha'})\\
-\frac{1}{z_i}c_i(\alpha)\sum_{\alpha'}c_{i-1}(\alpha')\sum_{\vec{\sigma}_{i-1}:D_{i,i-1}\ge d}g_{i,i-1}^{\alpha}g_{i-1,j}^{\alpha'}\prod_{k=1:k\neq j}^{i-2}(\sum_{\vec{\sigma}_k:D_{i,k},D_{i-1,k}\ge d}g_{i,k}^{\alpha}g_{i-1,k}^{\alpha'}),
\end{multline}
if $j=1,\dots,i-2$, and for $j=i-1$
\begin{multline}
\lambda_{i,i-1}^{\alpha}=c_i(\alpha)\sum_{\alpha'}c_i(\alpha')g_{i,i-1}^{\alpha'}\prod_{k=1}^{i-2}(\sum_{\vec{\sigma}_k:D_{i,k}\ge d}g_{i,k}^{\alpha}g_{i,k}^{\alpha'})\\
-\frac{1}{z_i}c_i(\alpha)\sum_{\alpha'}c_{i-1}(\alpha')\prod_{k=1}^{i-2}(\sum_{\vec{\sigma}_k:D_{i,k},D_{i-1,k}\ge d}g_{i,k}^{\alpha}g_{i-1,k}^{\alpha'}).
\end{multline}
The Lagrange multipliers $\lambda_i$ and $\lambda_{ij}^{\alpha}$ ensure that 
\begin{align}
\sum_{\alpha}c_i(\alpha) &=1,\\
\sum_{\vec{\sigma}_j:D_{i,j}\ge d}g_{i,j}^{\alpha}(\vec{\sigma}_j:\vec{\sigma}_i) &=1.
\end{align}

To be specific, we take 
\begin{align}
\sum_{\alpha}c_i(\alpha)[\dots]=\sum_{l_{j=1,\dots,i-1}=d}^nc_i(l_1,\dots,l_{i-1})\frac{1}{(i-1)!}\sum_{P}[\dots].
\end{align}
This means that the $i-1$ spheres are distributed in distances $l_1,\dots,l_{i-1}$, and to ensure the permutation symmetry we sum over all the $(i-1)!$ permutations with equal weights. More precisely, now the the conditional part of the BP messages read as follows
\begin{align}
q_{i}(h_{i}:\vec{\sigma}_i)=\sum_{l_{j=1,\dots,i-1}\ge d}c_i(l_1,\dots,l_{i-1})\frac{1}{(i-1)!}\sum_{P}\prod_{j=1}^{i-1}g_{i,j}^{l_{Pj}}(\vec{\sigma}_j:\vec{\sigma}_i),
\end{align}
where as before $h_{i}=\{\vec{\sigma}_{1:i-1}\}$, and $j'=Pj$ shows the effect of permutation $P$ on $j$.

By symmetry, the coefficients $c_i(l_1,\dots,l_{i-1})=c_i(\{l_{1:i-1}\})$ depend on the number of spheres at distance $l$.
As before, we assume that the $g_{i,j}^{l_{Pj}}$ depend on Hamming distances $D_{i,j}$ and consider a complete and normalized representation
\begin{align}
g_{i,j}^{l_{Pj}}(\vec{\sigma}_j:\vec{\sigma}_i)=\frac{1}{w(l_{Pj})}\delta_{l_{ij},l_{Pj}}.
\end{align}
This fixes the values of the $g_{i,j}$ variables in the cavity marginals.  
We also defined $w(l)=C(l:n)$ for the number of points at distance $l$ from an arbitrary point in the Hamming space.
In this way, the stationary equations are given by
\begin{multline}
\lambda_{i}=\frac{1}{(i-1)!(i-1)!}\sum_{\{l_{1:i-1}'\}}\sum_{P,P'}c_i(\{l_{1:i-1}'\})\prod_{j=1}^{i-1}\left(\frac{\delta_{l_{Pj},l_{P'j}'}}{w(l_{Pj})}\right)\\
-\frac{1}{z_i}\frac{1}{(i-1)!(i-2)!}\sum_{\{l_{1:i-2}'\}}\sum_{P,P'}c_{i-1}(\{l_{1:i-2}'\})\sum_{\vec{\sigma}_{i-1}:D_{i,i-1}\ge d}g_{i,i-1}^{l_{P(i-1)}}\prod_{j=1}^{i-2}\left(\frac{\delta_{l_{Pj},l_{P'j}'}}{w(l_{Pj})}\right),
\end{multline}
which can be simplified to
\begin{align}
\lambda_{i}=\frac{1}{\prod_{j=1}^{i-1}w(l_j)}c_i(\{l_{1:i-1}\})
-\frac{1}{z_i}\frac{1}{(i-1)}\sum_{k=1}^{i-1}\frac{1}{\prod_{j\ne k}w(l_j)}c_{i-1}(\{l_{1:i-1}\}\setminus l_k).
\end{align}
Or, in terms of the $c_i$ variables we have   
\begin{align}
c_i(\{l_{1:i-1}\})=\lambda_{i}\prod_{j=1}^{i-1}w(l_j)
+\frac{1}{z_i}\frac{1}{(i-1)}\sum_{k=1}^{i-1}w(l_k)c_{i-1}(\{l_{1:i-1}\}\setminus l_k).
\end{align}
On the other hand, from the normalization constraints 
\begin{align}
\sum_{l_1,\dots,l_{i-1}=d}^n c_i(\{l_{1:i-1}\}) &=1,\\
\frac{1}{2^n}\sum_{\vec{\sigma}_i,\{\vec{\sigma}_{1:i-1}\}}q_i(\{\vec{\sigma}_{1:i-1}\}:\vec{\sigma}_i) &=1,
\end{align}
one obtains
\begin{align}
\lambda_i=\frac{1}{(2^n-V_d)^{i-1}}(1-\frac{2^n-V_d}{z_i}),
\end{align}
and
\begin{align}
z_i=\sum_{l_1,\dots,l_{i-1}=d}^n c_{i-1}(\{l_{1:i-2}\})w(l_{i-1})\prod_{j=1}^{i-2}\left(\sum_{l'=d}^n\Omega(l',l_j:l_{i-1})/w(l_j) \right).
\end{align}
Finally, after some straightforward algebra, we obtain a recursive equation for the $c_i$ in terms of the previously defined quantities  
\begin{multline}
c_i(\{l_{1:i-1}\})=\prod_{j=1}^{i-1}\left(\frac{w(l_j)}{2^n-V_d}\right)\\
+\frac{\sum_{k=1}^{i-1}w(l_k)c_{i-1}(\{l_{1:i-1}\}\setminus l_k)/(i-1)-(2^n-V_d)\prod_{j=1}^{i-1}(w(l_j)/(2^n-V_d))}{\sum_{l_1',\dots,l_{i-1}'=d}^n c_{i-1}(\{l_{1:i-2}'\})w(l_{i-1}')\prod_{j=1}^{i-2}\left(\sum_{l'=d}^n\Omega(l',l_j':l_{i-1}')/w(l_j') \right)}.
\end{multline}
A reasonable initial condition for the above equation is
\begin{align}
c_2(l_1)=\frac{w(l_1)}{2^n-V_d}.
\end{align}
This makes the second term for computing $c_3(l_1,l_2)$ zero and results in
\begin{align}
c_3(l_1,l_2)=\frac{w(l_1)}{2^n-V_d}\frac{w(l_2)}{2^n-V_d}.
\end{align}
The same happens for other coefficients and we obtain
\begin{align}
c_i(\{l_{1:i-1}\})=\prod_{j=1}^{i-1}\left(\frac{w(l_j)}{2^n-V_d}\right),
\end{align}
which is equivalent to a uniform distribution of the sphere positions $\{\vec{\sigma}_{1:i-1}\}$.

From the above solutions we obtain the node contributions to the entropy Eq. (\ref{dsi-chain}),
\begin{multline}
e^{\Delta S_i}=\frac{1}{2^{n(i+2)}}\sum_{l_1,\dots,l_{i-2}=d}^n c_{i-1}(\{l_{1:i-2}\})\frac{1}{(i-2)!}\sum_P\\
\sum_{\vec{\sigma}_{i-1},\vec{\sigma}_i,\vec{\sigma}_{i+1}:D_{i,i-1},D_{i,i+1},D_{i-1,i+1}\ge d}\prod_{j=1}^{i-2}\left(\sum_{\vec{\sigma}_j:D_{j,i-1},D_{j,i},D_{j,i+1}\ge d} g_{i-1,j}^{l_{Pj}}(\vec{\sigma}_j:\vec{\sigma}_{i-1})\right), 
\end{multline}
which simplifies to
\begin{align}
e^{\Delta S_i}=\frac{1}{2^{n(i+1)}}\sum_{l_1,l_2,l_3=d}^n C(l_1:n)\Omega(l_2,l_3:l_1)\left(\frac{1}{2^n-V_d}\sum_{l_{i-1},l_i,l_{i+1}=d}^n Y(l_{i-1},l_i,l_{i+1}:l_1,l_2,l_3)\right)^{i-2}.
\end{align}
The edge contributions are obtained from Eq. (\ref{dsij-chain}), 
\begin{align}
e^{\Delta S_{i,i+1}}= \frac{1}{2^{n(i+2)}}\sum_{l_1,\dots,l_{i-1}=d}^n c_{i}(\{l_{1:i-1}\})\frac{1}{(i-1)!}\sum_P
\sum_{\vec{\sigma}_i,\vec{\sigma}_{i+1}:D_{i,i+1}\ge d}\prod_{j=1}^{i-1}\left(\sum_{\vec{\sigma}_j:D_{j,i},D_{j,i+1}\ge d} g_{i,j}^{l_{Pj}}(\vec{\sigma}_j:\vec{\sigma}_i)\right),
\end{align}
or, after simplification
\begin{align}
e^{\Delta S_{i,i+1}}=\frac{1}{2^{n(i+1)}}\sum_{l=d}^n C(l:n)\left(\frac{1}{2^n-V_d}\sum_{l_i,l_{i+1}=d}^n \Omega(l_i,l_{i+1}:l)\right)^{i-1}.
\end{align}
As expected, these are identical expressions to those obtained for the entropy using the naive approximation of the BP messages.

\subsection{Inhomogeneous solutions}\label{S25}
Let us break the translation symmetry and look for more general solutions to the BP equations $\mu_{i}(\vec{\sigma}_i,h_i)= p_i(\vec{\sigma}_i) q_i(h_i:\vec{\sigma}_i)$ where $p_i(\vec{\sigma}_i)$ is not necessarily $1/2^n$. 
Assuming again that the conditional part is given by
\begin{align}
q_{i}(h_{i}:\vec{\sigma}_i)=\sum_{\alpha}c_i(\alpha)\prod_{j=1}^{i-1}\left( \mathbb{I}(D_{i,j}\ge d)g_{i,j}^{\alpha}(\vec{\sigma}_j:\vec{\sigma}_i) \right),
\end{align}
we minimize the resulted error in the BP equations
\begin{multline}\label{Lg-inh}
\mathcal{L}[p_i(\vec{\sigma}_i),\{c_i^{\alpha},g_{i,j=1,\dots,i-1}^{\alpha}\}]=\sum_{\vec{\sigma}_i,\{\vec{\sigma}_{j=1,\dots,i-1}:D_{i,j}\ge d\}}\\
\left( p_i(\vec{\sigma}_i)\sum_{\alpha}c_i(\alpha)\prod_{j=1}^{i-1}g_{i,j}^{\alpha}-\frac{1}{z_i}p_{i-1}(\vec{\sigma}_{i-1})\sum_{\alpha}c_{i-1}(\alpha)\prod_{j=1}^{i-2}\mathbb{I}(D_{i-1,j}\ge d)g_{i-1,j}^{\alpha}\right)^2.
\end{multline}
The main equations and details of computations for this general case are given in Appendix \ref{app2}.

In the following, we continue with the last expression for the $q_i(h_i:\vec{\sigma}_i)$ in the previous subsection which respects the permutation symmetry, that is
\begin{align}
q_{i}(h_{i}:\vec{\sigma}_i)=\sum_{l_{j=1,\dots,i-1}\ge d}c_i(l_1,\dots,l_{i-1})\frac{1}{(i-1)!}\sum_{P}\prod_{j=1}^{i-1}g_{i,j}^{l_{Pj}}(\vec{\sigma}_j:\vec{\sigma}_i).
\end{align}
As before $g_{i,j}^{l_{Pj}}(\vec{\sigma}_j:\vec{\sigma}_i)=\frac{1}{w(l_{Pj})}\delta_{l_{ij},l_{Pj}}$ is fixed and we have only two stationary equations for the $p_i(\vec{\sigma}_i)$ and $c_i(\{l_{1:i-1}\})$. Derivation of the error function Eq. (\ref{Lg-inh}) with respect to the $p_i(\vec{\sigma}_i)$ in presence of the normalization constraints leads to
\begin{multline}\label{inh-pi}
p_i(\vec{\sigma}_i)=\gamma_i'
+\frac{1}{\sum_{\{l_{1:i-1}\ge d\}}c_i^2(\{l_{1:i-1}\})/\prod_{j=1}^{i-1}w(l_j)}\sum_{\{l_{1:i-1}\ge d\}}\frac{c_i(\{l_{1:i-1}\})}{\prod_{j=1}^{i-1}w(l_j)}\\ 
\times \frac{1}{(i-1)!}\sum_{P}\frac{1}{z_i}\sum_{\{l_{1:i-2}'\ge d\}}c_{i-1}(\{l_{1:i-2}'\}) \sum_{\vec{\sigma}_{i-1}:D_{i,i-1}=l_{P(i-1)}}p_{i-1}(\vec{\sigma}_{i-1})
\prod_{j=1}^{i-2}\left( \frac{\Omega(l_{Pj},l_{j}':l_{P(i-1)})}{w(l_{j}')} \right).
\end{multline}
The stationary equations for the $c_i$ read as follows 
\begin{multline}\label{inh-cl}
c_i(\{l_{1:i-1}\})=\lambda_i'\prod_{j=1}^{i-1}w(l_j)
+\frac{1}{\sum_{\vec{\sigma}_i}p_i^2(\vec{\sigma}_i)}\sum_{\vec{\sigma}_i}p_i(\vec{\sigma}_i)\\ \times
\frac{1}{(i-1)!}\sum_{P}\frac{1}{z_i}\sum_{\{l_{1:i-2}'\ge d\}}c_{i-1}(\{l_{1:i-2}'\})\sum_{\vec{\sigma}_{i-1}:D_{i,i-1}=l_{P(i-1)}}p_{i-1}(\vec{\sigma}_{i-1})\prod_{j=1}^{i-2}\left( \frac{\Omega(l_{Pj},l_{j}':l_{P(i-1)})}{w(l_{j}')} \right).
\end{multline}
The normalization factor $z_i$ is 
\begin{align}
z_i=\sum_{\{l_{1:i-2}'\ge d\}}^n c_{i-1}(\{l_{1:i-2}'\})\sum_{\vec{\sigma}_i,\vec{\sigma}_{i-1}:l_{i,i-1}\ge d}p_{i-1}(\vec{\sigma}_{i-1})\prod_{j=1}^{i-2}\left(\sum_{l=d}^n\frac{\Omega(l,l_j':l_{i,i-1})}{w(l_j')} \right),
\end{align}
and the coefficients $\gamma_i'$ and $\lambda_i'$ are Lagrange multipliers.

Note that the first terms of the two equations are expected from a homogeneous solution and in both equations we have a term like
\begin{multline}
Q(\vec{\sigma}_i,\{l_{1:i-1}\})\equiv \frac{1}{(i-1)!}\sum_{P}\frac{1}{z_i}\sum_{\{l_{1:i-2}'\ge d\}}c_{i-1}(\{l_{1:i-2}'\})
\\ \times
\sum_{\vec{\sigma}_{i-1}:D_{i,i-1}=l_{P(i-1)}}p_{i-1}(\vec{\sigma}_{i-1})\prod_{j=1}^{i-2}\left( \frac{\Omega(l_{Pj},l_{j}':l_{P(i-1)})}{w(l_{j}')} \right).
\end{multline}
Normalization conditions give the Lagrange multipliers
\begin{align}
\gamma_i' &=\frac{1}{2^n}\left(1-
\frac{1}{\sum_{\{l_{1:i-1}\ge d\}}c_i^2(\{l_{1:i-1}\})/\prod_{j=1}^{i-1}w(l_j)}\sum_{\{l_{1:i-1}\ge d\}}\frac{c_i(\{l_{1:i-1}\})}{\prod_{j=1}^{i-1}w(l_j)}\sum_{\vec{\sigma}_i} Q(\vec{\sigma}_i,\{l_{1:i-1}\}) \right),\label{gi} \\
\lambda_i' &=\frac{1}{(2^n-V_d)^{i-1}}\left(1-\frac{1}{\sum_{\vec{\sigma}_i}p_i^2(\vec{\sigma}_i)}\sum_{\vec{\sigma}_i}p_i(\vec{\sigma}_i)\sum_{\{l_{1:i-1}\ge d\}}Q(\vec{\sigma}_i,\{l_{1:i-1}\}) \right).\label{li}
\end{align}
We rewrite the above equations for the $p_i$ and $c_i$ in a more compact form   
\begin{align}\label{inh}
p_i(\vec{\sigma}_i) &=\frac{1}{2^n}+\Delta p_i(\vec{\sigma}_i),\\
c_i(\{l_{1:i-1}\}) &=\prod_{j=1}^{i-1}\left(\frac{w(l_j)}{2^n-V_d}\right)+\Delta c_i(\{l_{1:i-1}\}),
\end{align}
with deviations $\Delta p_i,\Delta c_i$ from the liquid solution defined by
\begin{multline}\label{inh-dp}
\Delta p_i(\vec{\sigma}_i)\equiv \frac{1}{\sum_{\{l_{1:i-1}\ge d\}}c_i^2(\{l_{1:i-1}\})/\prod_{j=1}^{i-1}w(l_j)} \\ \times
\sum_{\{l_{1:i-1}\ge d\}}\frac{c_i(\{l_{1:i-1}\})}{\prod_{j=1}^{i-1}w(l_j)}\left(Q(\vec{\sigma}_i,\{l_{1:i-1}\})-\frac{1}{2^n}\sum_{\vec{\sigma}_i'}Q(\vec{\sigma}_i',\{l_{1:i-1}\})\right),
\end{multline}
and
\begin{multline}\label{inh-dc}
\Delta c_i(\{l_{1:i-1}\})\equiv \frac{1}{\sum_{\vec{\sigma}_i}p_i^2(\vec{\sigma}_i)}\sum_{\vec{\sigma}_i}p_i(\vec{\sigma}_i)
\\ \times
\left(Q(\vec{\sigma}_i,\{l_{1:i-1}\})-\prod_{j=1}^{i-1}\left(\frac{w(l_j)}{2^n-V_d}\right)\sum_{\{l_{1:i-1}'\ge d\}}Q(\vec{\sigma}_i,\{l_{1:i-1}'\})\right).
\end{multline}

Let us rewrite the equations for the $p_i$ and $c_i$ as 
\begin{align}
[p_i(\vec{\sigma}_i)-\gamma_i']\left( \sum_{\{l_{1:i-1}\ge d\}}\frac{c_i^2(\{l_{1:i-1}\})}{\prod_{j=1}^{i-1}w(l_j)} \right)&=\sum_{\{l_{1:i-1}\ge d\}}\frac{c_i(\{l_{1:i-1}\})}{\prod_{j=1}^{i-1}w(l_j)}Q(\vec{\sigma}_i,\{l_j\}),\\
[c_i(\{l_{1:i-1}\})-\lambda_i'\prod_{j=1}^{i-1}w(l_j)]\left(\sum_{\vec{\sigma}_i}p_i^2(\vec{\sigma}_i)\right)&=\sum_{\vec{\sigma}_i}p_i(\vec{\sigma}_i)Q(\vec{\sigma}_i,\{l_{1:i-1}\}).
\end{align}
Now, we multiply the first equation by $p_i(\vec{\sigma}_i)$ and sum over $\vec{\sigma}_i$. Using the normalization condition and the second equation we get
\begin{multline}
[\sum_{\vec{\sigma}_i}p_i^2(\vec{\sigma}_i)-\gamma_i']\left( \sum_{\{l_{1:i-1}\ge d\}}\frac{c_i^2(\{l_{1:i-1}\})}{\prod_{j=1}^{i-1}w(l_j)} \right)\\
=\left(\sum_{\vec{\sigma}_i}p_i^2(\vec{\sigma}_i)\right)\sum_{\{l_{1:i-1}\ge d\}}\frac{c_i(\{l_{1:i-1}\})}{\prod_{j=1}^{i-1}w(l_j)}[c_i(\{l_{1:i-1}\})-\lambda_i'\prod_{j=1}^{i-1}w(l_j)].
\end{multline}
Simplifying the equation results in
\begin{align}\label{gili}
\gamma_i'\left( \sum_{\{l_{1:i-1}\ge d\}}\frac{c_i^2(\{l_{1:i-1}\})}{\prod_{j=1}^{i-1}w(l_j)} \right)
=\lambda_i'\left(\sum_{\vec{\sigma}_i}p_i^2(\vec{\sigma}_i)\right),
\end{align}
where we also used normalization condition $\sum_{\{l_{1:i-1}\ge d\}}c_i(\{l_{1:i-1}\})=1$.

For the liquid case
\begin{align}
p_i(\vec{\sigma}_i) &=\frac{1}{2^n},\\ 
c_i(\{l_{1:i-1}\}) &=\prod_{j=1}^{i-1}\left(\frac{w(l_j)}{(2^n-V_d)}\right),
\end{align}
a consistent solution to the equations is given by
\begin{align}\label{hQ}
Q(\vec{\sigma}_i,\{l_{1:i-1}\}=\frac{(i-1)}{2^n}\prod_{j=1}^{i-1}\left(\frac{w(l_j)}{2^n-V_d}\right),
\end{align}
with $\Delta p_i=\Delta c_i=0$. In addition, for Eqs. \ref{gi}, \ref{li}, and \ref{gili} we obtain
\begin{align}
\gamma_i' &=\frac{1}{2^n}\left(1-\sum_{\vec{\sigma}_i,\{l_{1:i-1}\ge d\}}Q(\vec{\sigma}_i,\{l_{1:i-1}\})\right),\\
\lambda_i' &=\frac{1}{(2^n-V_d)^{i-1}}\left(1-\sum_{\vec{\sigma}_i,\{l_{1:i-1}\ge d\}}Q(\vec{\sigma}_i,\{l_{1:i-1}\})\right),\\
\frac{\lambda_i'}{2^n} &=\frac{\gamma_i'}{(2^n-V_d)^{i-1}},
\end{align}
which are satisfied by the liquid solution.

\begin{figure}
\includegraphics[width=14cm]{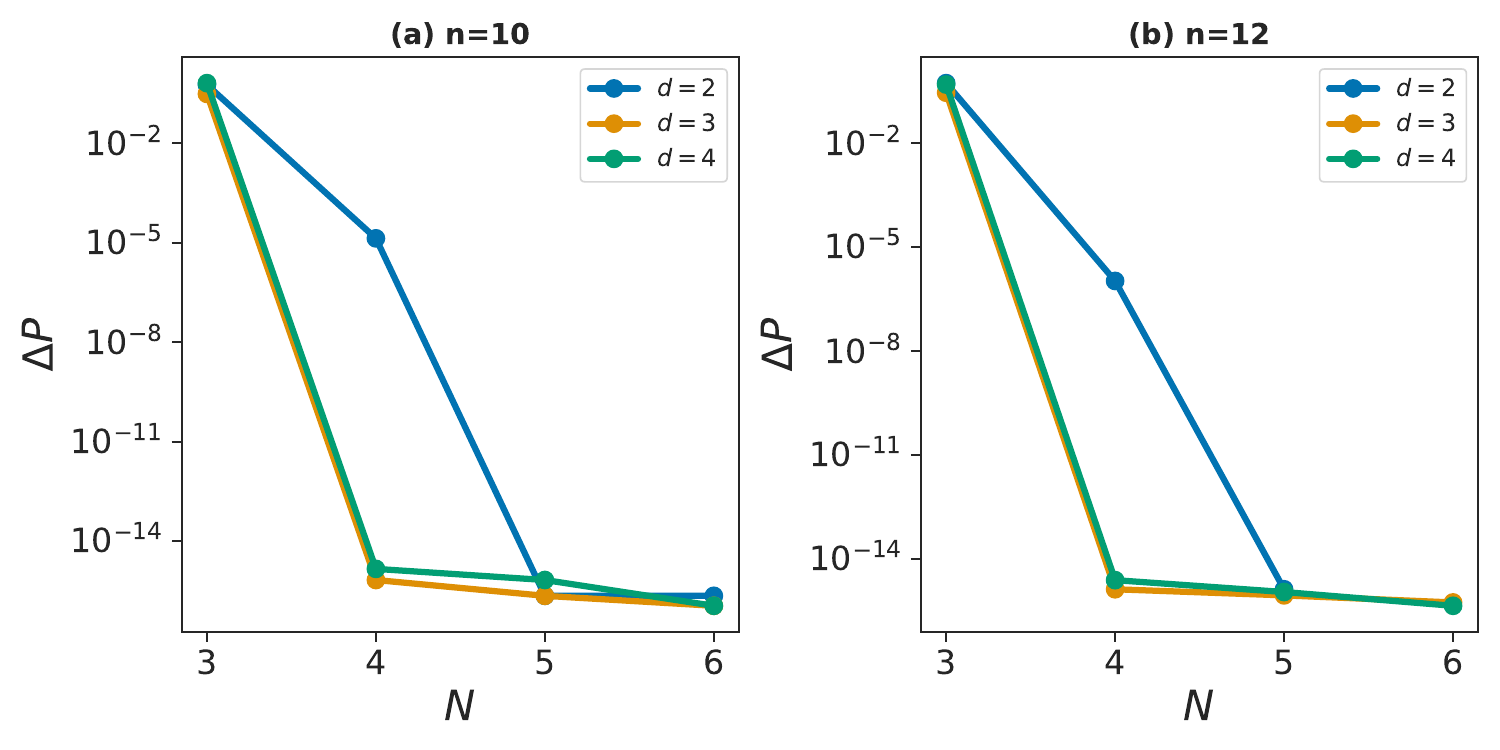} 
\caption{Deviation from the uniform solution when we start from one sphere at origin. The probabilities $p_i(\vec{\sigma}_i)$ are computed numerically from the BP equations with the inhomogeneous ansatz for the cavity marginals. We assume that the other probabilities $c_{i}(\{l_j\})$ are concentrated on a single configuration $\{l_j^*\}$. The reported deviation $\Delta P=\sum_{\vec{\sigma}}|p_i(\vec{\sigma})-\frac{1}{2^n}|$ is maximized over all possible configurations of $\{l_j^*\}$.}\label{fig3}
\end{figure}

On the other side, we may consider a frozen solution, where the probability distributions are concentrated on a single configuration, for instance, 
\begin{align}
p_i(\vec{\sigma}_i) &=\delta_{\vec{\sigma}_i,\vec{\sigma}_i^*},\\ 
c_i(\{l_{1:i-1}\}) &=\delta_{\{l_{1:i-1}\},\{l_{1:i-1}^*\}}.
\end{align}
In this case, from Eqs. \ref{gi}, \ref{li}, and \ref{gili} we find
\begin{align}
\gamma_i' &=\frac{1}{2^n}\left(1-\sum_{\vec{\sigma}_i}Q(\vec{\sigma}_i,\{l_{1:i-1}^*\})\right),\\
\lambda_i' &=\frac{1}{(2^n-V_d)^{i-1}}\left(1-\sum_{\{l_{1:i-1}\ge d\}}Q(\vec{\sigma}_i^*,\{l_{1:i-1}\})\right),\\
\lambda_i' &=\frac{\gamma_i'}{\prod_{j=1}^{i-1}w(l_j^*)}.
\end{align}
This results in a consistency equation which should be satisfied by the $(\vec{\sigma}_i^*,\{l_{1:i-1}^*\})$ given the previous assignments for $j=1,\cdots,i-1$,
\begin{align}
\frac{1}{2^n}\left(1-\sum_{\vec{\sigma}_i}Q(\vec{\sigma}_i,\{l_{1:i-1}^*\})\right)=
\frac{\prod_{j=1}^{i-1}w(l_j^*)}{(2^n-V_d)^{i-1}}\left(1-\sum_{\{l_{1:i-1}\ge d\}}Q(\vec{\sigma}_i^*,\{l_{1:i-1}\})\right).
\end{align}
Starting from an initial condition one can try to satisfy the above equation by iteration to find a configuration $(\vec{\sigma}_i^*,\{l_{1:i-1}^*\})$ for sphere $i$. An initial condition starting from $i=2$ could be 
\begin{align}
p_2(\vec{\sigma}_2)=\delta_{\vec{\sigma}_2,\vec{0}},\\
c_2(l_1)=\delta_{l_1,d}.
\end{align}
Let us assume that the probabilities $c_{i}(\{l_j\})$ are concentrated on a single configuration $\{l_j^*\}$. Then we compute the $p_i(\vec{\sigma}_i)$ from Eq. \ref{inh-pi} using the above initial condition. The deviation $\Delta P=\sum_{\vec{\sigma}}|p_i(\vec{\sigma})-\frac{1}{2^n}|$ is then maximized over all possible configurations of $\{l_j^*\}$. 
Figure \ref{fig3} shows that the above process converges quickly to a uniform solution $p_i(\vec{\sigma}_i)=1/2^n$ as the number of spheres $N$ increases. The minimum of deviation is zero for all the points which are reported in the figure.   

\section{Conclusion}\label{S3}
We used an exact representation of the entropy of a system of hard spheres in the Hamming space to investigate the asymptotic behavior of the maximum packing density of the spheres. The method is based on a decomposition of the interaction graph into a tree structure and the induced loopy interactions, which their effects are taken into account by passing some (internal) messages along the tree interaction graph. The solutions to the Bethe equations for reasonable approximations of the BP cavity marginals are asymptotically consistent with the Gilbert-Varshamov lower bound for the packing density, but we can not exclude the possibility of solutions of higher densities.       

We considered an ansatz of BP marginals which are represented by a linear superposition of factorized probability distributions which respect the permutation symmetry of the problem. It would be interesting to try other classes of tractable cavity marginals as trial BP messages which are constrained by satisfying the exact BP equations. For instance, another possibility is working with simpler internal messages $h_{i\to j}$ instead of approximating the BP cavity marginals. Given an ansatz of reasonable BP probability marginals, one can also try to maximize the Bethe entropy to find an upper bound within a specific class of solutions to the BP equations.

\acknowledgments
We would like to thank Parsa Rangriz and Francesco Zamponi for helpful discussions and useful comments. This work was performed using the Academic Leiden Interdisciplinary Cluster Environment (ALICE), the compute resources provided by Leiden University.

\appendix

\section{The naive approximation of the BP messages: Asymptotic entropy}\label{app1}
The BP probability marginals $\mu_{i}(\vec{\sigma}_i,h_i)$ in general depend on the position probability of sphere $i$, $p_i(\vec{\sigma}_i)$ and a conditional probability $q_i(h_i:\vec{\sigma}_i)$, 
\begin{align}
\mu_{i}(\vec{\sigma}_i,h_i)= p_i(\vec{\sigma}_i) q_i(h_i:\vec{\sigma}_i).
\end{align}
The $h_i=\{\vec{\sigma}_{j=1,\dots,i-1}\}$ represent the positions of all spheres $j=1,\cdots,i-1$.
By symmetry we take a uniform distribution for $\vec{\sigma}_i$,
\begin{align}
p_{i}(\vec{\sigma}_i)=\frac{1}{2^n}.
\end{align}
We also assume that the probability distribution of the other spheres is factorized and uniform
\begin{align}
q_{i}(h_{i}:\vec{\sigma}_i)=\prod_{j=1}^{i-1}\frac{\mathbb{I}(D_{i,j}\ge d)}{2^n-V_d}.
\end{align}

In this way, for the nodes' contribution to the entropy we have
\begin{multline}
e^{\Delta S_i}=\frac{1}{2^{n+ni}}\sum_{l_1,l_2,l_{12}=d}^n C(l_{12}:n)
\Omega(l_1,l_2:l_{12})\\
\times \left(1-\frac{2V_d-O_{01}(l_1)-O_{02}(l_2)-O_{12}(l_{12})+2O_{012}(l_1,l_2,l_{12})}{2^n-V_d}\right)^{i-2}.
\end{multline}
The edges' contribution in the entropy are
\begin{align}
e^{\Delta S_{i,i+1}}= \frac{1}{2^{n+ni}}\sum_{l=d}^n C(l:n)
\left(1-\frac{V_d-O_{01}(l)}{2^n-V_d}\right)^{i-1}.
\end{align}
Here $O_{ij}(l)$ is the overlap of two spheres of radius $d$ at distance $l$. And $O_{ijk}(l_i,l_j,l_{ij})$ is the overlap of three spheres of radius $d$ when $(i,j)$ have distance $l_{ij}$ and the other sphere is at distances $l_i,l_j$ form $(i,j)$.
The function $\Omega(l_i,l_j:l_{ij})$ is the number of possible points for the third sphere given the distances,
\begin{align}
\Omega(l_i,l_j:l_{ij})= C(\frac{l_i-l_j+l_{ij}}{2}:l_{ij})C(\frac{l_i+l_j-l_{ij}}{2}:n-l_{ij}).
\end{align}
Thus for the overlaps we get
\begin{align}
O_{ij}(l_{ij})=\sum_{l_i,l_j=0}^d\Omega(l_i,l_j:l_{ij}).
\end{align}
Let us define $Y(l_i,l_j,l_k:l_{ij},l_{ik},l_{jk})$ as the number of points at distances $l_i,l_j,l_k$ from spheres $i,j,k$ with distances $l_{ij},l_{ik},l_{jk}$,
\begin{multline}
Y(l_i,l_j,l_k:l_{ij},l_{ik},l_{jk})\\
=\sum_{x_{00}=0}^n C(x_{11}:x)C(x_{01}:l_{ik}-x)C(x_{00}:n-l_{ik}-l_{jk}+x)C(x_{10}:l_{jk}-x)
\end{multline}
with
\begin{align}
2x &=l_{ik}+l_{jk}-l_{ij}, \\
2x_{01} &=l_j+l_k-l_{jk}-2x_{00}, \\
2x_{10} &=l_i+l_k-l_{ik}-2x_{00}, \\
2x_{11} &=l_{ik}+l_{jk}-l_i-l_j+2x_{00}. \\
\end{align}
Then
\begin{align}
O_{ijk}(l_i,l_j,l_{ij})=\sum_{l_i,l_j,l_k=0}^dY(l_i,l_j,l_k:l_{ij},l_{ik},l_{jk}).
\end{align}

Let us take the limit $n,d\to \infty$ with $\delta=d/n$ finite. All distances are scaled with $n$, for instance $r_{ij}=l_{ij}/n$. Then $v_d=V_d/2^n=e^{-n[\ln(2)-H(\delta)]}$ with $H(\delta)=-\delta \ln(\delta)-(1-\delta) \ln(1-\delta)$ is an exponentially small quantity. The node's contribution in the entropy is given by  
\begin{multline}
\Delta S_{node}=\sum_{i=1}^N\Delta S_i=\simeq nN\int_0^1 dx[H^*(x)+H_1^*(x)+H_2^*(x)\\
-x\tilde{v}_d(2-o_{01}^*(x)-o_{02}^*(x)-o_{12}^*(x)+2o_{012}^*(x))]
-nN\frac{\ln(2)}{2},
\end{multline}
after approximating $\sum_i \approx N\int_0^1 dx$.
The overlaps are scaled $o=O/V_d$ and $\tilde{v}_d=Nv_d/n$. The star means the above quantities are computed at $r^*(x),r_1^*(x),r_2^*(x)$ which
\begin{multline}
r^*(x),r_1^*(x),r_2^*(x)=
\arg\max_{r,r_1,r_2 \in (\delta,1)}[ H(r)+H_1(r_1,r_2,r)+H_2(r_1,r_2,r)\\
-x\tilde{v}_d(2-o_{01}(r_1)-o_{02}(r_2)-o_{12}(r)+2o_{012}(r_1,r_2,r))].
\end{multline}
Here
\begin{align}
H_1(r_1,r_2,r)=r\ln r-\frac{r_1-r_2+r}{2}\ln\frac{r_1-r_2+r}{2}- \frac{r_2-r_1+r}{2}\ln\frac{r_2-r_1+r}{2},
\end{align}
and
\begin{multline}
H_2(r_1,r_2,r)=(1-r)\ln(1-r)\\
-\frac{r_1+r_2-r}{2}\ln\frac{r_1+r_2-r}{2}- (1-\frac{r_1+r_2+r}{2})\ln(1-\frac{r_1+r_2+r}{2}).
\end{multline}

\begin{figure}
\includegraphics[width=10cm]{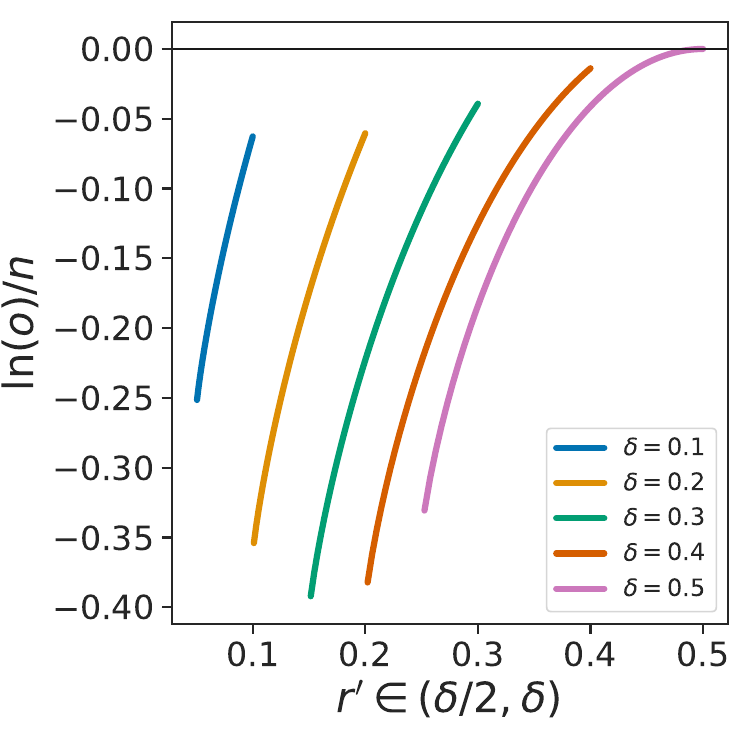} 
\caption{The overlap $o=O/V_d$ of two spheres of scaled diameter $\delta$ at distance $r'$. Numerical computation of this quantity shows that it is exponentially small for $\delta <1/2$.}\label{fig4}
\end{figure}

The overlaps are exponentially small here. Consider for instance the maximum overlap $o_{ij}(r)$ at $r=\delta$,
\begin{align}
o_{ij}(\delta)=\frac{O_{ij}}{V_d}\simeq e^{n[H_1(r',r',\delta)+H_2(r',r',\delta)-H(\delta)]}.
\end{align}
By symmetry, the maximum is for $r_1=r_2=r'$. A plot of the exponent $H_1(r',r',\delta)+H_2(r',r',\delta)-H(\delta)$ as a function of $r'$ in Fig. \ref{fig4} shows that this quantity is always negative for $\delta<1/2$.

Since the overlaps are exponentially small, they can be ignored to get
\begin{align}
\Delta S_{node}\simeq nN\left(H(\frac{1}{2})+H_1(\frac{1}{2},\frac{1}{2},\frac{1}{2})+H_2(\frac{1}{2},\frac{1}{2},\frac{1}{2})-\frac{1}{2}(2\tilde{v}_d+\ln(2)) \right),
\end{align}

The link's contribution in the entropy is given by  
\begin{align}
\Delta S_{link}=\sum_{i=1}^{N-1}\Delta S_{i,i+1}\simeq nN\left(\int_0^1 dx[H^*(x)-x\tilde{v}_d(1-o_{ij}^*(x))]-\frac{\ln(2)}{2} \right).
\end{align}
The star means the above quantities are computed at $r^*(x)$ which
\begin{align}
r^*(x)=\arg\max_{r\in (\delta,1)}\left( H(r)-x\tilde{v}_d(1-o_{ij}(r))\right).
\end{align}
The exponentially small $o_{ij}$ can be ignored to get
\begin{align}
\Delta S_{link}\simeq nN\left(H(\frac{1}{2})-\frac{1}{2}(\tilde{v}_d+\ln(2)) \right),
\end{align}

Putting all together the entropy reads
\begin{align}
\Delta S_{node}\simeq nN\left(\frac{3}{2}\ln 2-\tilde{v}_d\right),\\
\Delta S_{link}\simeq nN\frac{1}{2}\left(\ln 2-\tilde{v}_d\right),\\
S^{naive}=\Delta S_{node}-\Delta S_{link}\simeq nN\left(\ln 2-\frac{1}{2}\tilde{v}_d\right).
\end{align}
Therefore, the entropy vanishes at
\begin{align}
N_{max}^{naive}=(2\ln 2)\frac{n}{v_d},
\end{align}
at the same point that the Bethe entropy in Eq. \ref{Sbethe} goes to zero.

\section{The BP equations for inhomogeneous solutions}\label{app2}
Let us break the translation symmetry and look for more general solutions to the BP equations $\mu_{i}(\vec{\sigma}_i,h_i)= p_i(\vec{\sigma}_i) q_i(h_i:\vec{\sigma}_i)$ where $p_i(\vec{\sigma}_i)$ is not necessarily $1/2^n$. 
Assuming that
\begin{align}
q_{i}(h_{i}:\vec{\sigma}_i)=\sum_{\alpha}c_i(\alpha)\prod_{j=1}^{i-1}\left( \mathbb{I}(D_{i,j}\ge d)g_{i,j}^{\alpha}(\vec{\sigma}_j:\vec{\sigma}_i) \right),
\end{align}
we minimize the expected error in the BP equations
\begin{multline}
\mathcal{L}[p_i(\vec{\sigma}_i),\{c_i^{\alpha},g_{i,j=1,\dots,i-1}^{\alpha}\}]=\sum_{\vec{\sigma}_i,\{\vec{\sigma}_{j=1,\dots,i-1}:D_{i,j}\ge d\}}\\
\left( p_i(\vec{\sigma}_i)\sum_{\alpha}c_i(\alpha)\prod_{j=1}^{i-1}g_{i,j}^{\alpha}-\frac{1}{z_i}p_{i-1}(\vec{\sigma}_{i-1})\sum_{\alpha}c_{i-1}(\alpha)\prod_{j=1}^{i-2}\mathbb{I}(D_{i-1,j}\ge d)g_{i-1,j}^{\alpha}\right)^2,
\end{multline}
The above function should be minimized with respect to the variables $p_i(\vec{\sigma}_i),\{c_i^{\alpha},g_{i,j=1,\dots,i-1}^{\alpha}\}$ conditioned to the following normalization constraints:
\begin{align}
\sum_{\vec{\sigma}_i}p_i(\vec{\sigma}_i) &=1,\\ 
\sum_{\alpha}c_i(\alpha) &=1,\\
\sum_{\vec{\sigma}_j:D_{i,j}\ge d}g_{i,j}^{\alpha}(\vec{\sigma}_j:\vec{\sigma}_i) &=1.
\end{align}

The stationary equations with respect to the variables $p_i(\vec{\sigma}_i),\{c_i^{\alpha},g_{i,j=1,\dots,i-1}^{\alpha}\}$ read as follows 
\begin{multline}
\gamma_i=p_i(\vec{\sigma}_i)\sum_{\alpha,\alpha'}c_i(\alpha)c_i(\alpha')\prod_{j=1}^{i-1}(\sum_{\vec{\sigma}_j:D_{i,j}\ge d}g_{i,j}^{\alpha}g_{i,j}^{\alpha'})\\
-\frac{1}{z_i}\sum_{\alpha,\alpha'}c_i(\alpha)c_{i-1}(\alpha')\sum_{\vec{\sigma}_{i-1}:D_{i,i-1}\ge d}p_{i-1}(\vec{\sigma}_{i-1})g_{i,i-1}^{\alpha}\prod_{j=1}^{i-2}(\sum_{\vec{\sigma}_j:D_{i,j},D_{i-1,j}\ge d}g_{i,j}^{\alpha}g_{i-1,j}^{\alpha'}),
\end{multline}

\begin{multline}
\lambda_i=\sum_{\vec{\sigma}_i}p_i^2(\vec{\sigma}_i)\sum_{\alpha'}c_i(\alpha')\prod_{j=1}^{i-1}(\sum_{\vec{\sigma}_j:D_{i,j}\ge d}g_{i,j}^{\alpha}g_{i,j}^{\alpha'})\\
-\frac{1}{z_i}\sum_{\vec{\sigma}_i}p_i(\vec{\sigma}_i)\sum_{\alpha'}c_{i-1}(\alpha')\sum_{\vec{\sigma}_{i-1}:D_{i,i-1}\ge d}p_{i-1}(\vec{\sigma}_{i-1})g_{i,i-1}^{\alpha}\prod_{j=1}^{i-2}(\sum_{\vec{\sigma}_j:D_{i,j},D_{i-1,j}\ge d}g_{i,j}^{\alpha}g_{i-1,j}^{\alpha'}),
\end{multline}
and
\begin{multline}
\lambda_{ij}^{\alpha}=p_i^2(\vec{\sigma}_i)c_i(\alpha)\sum_{\alpha'}c_i(\alpha')g_{i,j}^{\alpha'}\prod_{k=1:k\neq j}^{i-1}(\sum_{\vec{\sigma}_k:D_{i,k}\ge d}g_{i,k}^{\alpha}g_{i,k}^{\alpha'})\\
-\frac{1}{z_i}p_i(\vec{\sigma}_i)c_i(\alpha)\sum_{\alpha'}c_{i-1}(\alpha')\sum_{\vec{\sigma}_{i-1}:D_{i,i-1}\ge d}p_{i-1}(\vec{\sigma}_{i-1})g_{i,i-1}^{\alpha}g_{i-1,j}^{\alpha'}\prod_{k=1:k\neq j}^{i-2}(\sum_{\vec{\sigma}_k:D_{i,k},D_{i-1,k}\ge d}g_{i,k}^{\alpha}g_{i-1,k}^{\alpha'}),
\end{multline}
if $j=1,\dots,i-2$, and for $j=i-1$
\begin{multline}
\lambda_{i,i-1}^{\alpha}=p_i^2(\vec{\sigma}_i)c_i(\alpha)\sum_{\alpha'}c_i(\alpha')g_{i,i-1}^{\alpha'}\prod_{k=1}^{i-2}(\sum_{\vec{\sigma}_k:D_{i,k}\ge d}g_{i,k}^{\alpha}g_{i,k}^{\alpha'})\\
-\frac{1}{z_i}p_i(\vec{\sigma}_i)p_{i-1}(\vec{\sigma}_{i-1})c_i(\alpha)\sum_{\alpha'}c_{i-1}(\alpha')\prod_{k=1}^{i-2}(\sum_{\vec{\sigma}_k:D_{i,k},D_{i-1,k}\ge d}g_{i,k}^{\alpha}g_{i-1,k}^{\alpha'}).
\end{multline}
The Lagrange multipliers $\gamma_i, \lambda_i$ and $\lambda_{ij}^{\alpha}$ are to ensure the normalization constraints.

\end{document}